\documentclass[twocolumn,aps,pre,showpacs]{revtex4}
\usepackage[latin1]{inputenc}
\usepackage{graphicx}
\usepackage{latexsym}
\usepackage{amssymb}
\usepackage{amsmath}

\def\d{{\rm d}}

\def\x{{\bf x}}

\def\k{{\bf k}}

\def\N{{\bf N}}

\def\bzeta{{\boldsymbol\zeta}}

\def\aPP{a_{\scriptscriptstyle{PP}}}
\def\aPZ{a_{\scriptscriptstyle{PZ}}}
\def\aZP{a_{\scriptscriptstyle{ZP}}}
\def\aZZ{a_{\scriptscriptstyle{ZZ}}}
\def\aPPi{a_{{\scriptscriptstyle{PP}},i}}
\def\aPZi{a_{{\scriptscriptstyle{PZ}},i}}
\def\aZPi{a_{{\scriptscriptstyle{ZP}},i}}
\def\aZZi{a_{{\scriptscriptstyle{ZZ}},i}}

\def\beq{\begin{eqnarray}}
\def\eeq{\end{eqnarray}}
\begin{document}

\title{Effect of demographic noise in a phytoplankton-zooplankton
model of bloom dynamics}
\author{Piero Olla}
\affiliation{ISAC-CNR and INFN, Sez. Cagliari, I--09042 Monserrato, Italy.}
\date{\today}

\begin{abstract}
An extension of the Truscott-Brindley model (Bull. Math. Biol. {\bf 56}, 981 (1994))
is derived to account for the effect of demographic fluctuations. In the presence
of seasonal forcing, and sufficiently shallow water conditions, the fluctuations induced
by the discreteness of the zooplankton component appear sufficient to cause switching
between the bloom and no-bloom cycle predicted at the mean-field level by the model.
The destabilization persists in the thermodynamic limit of a
water basin infinitely extended in the horizontal direction.
\end{abstract}

\maketitle
\section{Introduction}
Phytoplankton provide the basis of the food chain in the ocean.
They are constituted for the large part by microscopic algae
and their abundance is such that they are responsible for roughly one half of the 
total photosynthesis going on in the planet \cite{field98}. 

Although phytoplankton are present in virtually all of the world hydrosphere
(more precisely, the top $100\,$m layer of the water column, called the euphotic layer,
where there is enough light for photosynthesis),
their distribution in space and time is far from uniform \cite{franks97}.  Space patchiness
is observed at scales that go from that of the individual to several hundreds
kilometers \cite{blackburn98,martin05}. 
The spatial patterns, revealed e.g. by remote sensing, include filaments,
fronts and more irregular shapes, suggesting that turbulent transport 
by sea currents may play an important role \cite{reigada03,metcalfe04,levy08,bracco09,mckiver11}.

Variability in time, on the other hand, is characterized by sporadic bloom events in which the  
plankton density in extended regions can grow by up to three orders of magnitude 
in few days \cite{franks97a}. These are seasonal events that may or may not recur annually, and
are superimposed onto the weaker day-night and seasonal cycles. Their importance is utmost
for several reasons. Depending on the species (e.g. diatoms), due to their abundance, bloom events
can have consequences even at the level of the global biogeochemical cycle
\cite{siegenhalter93}. Bloom events
involving other species (e.g. dinoflagellates) can have harmful effects on water
quality and fishery \cite{smayda97,yentsch08}. 
Being able to predict the onset of algal blooms is clearly
an issue of great practical importance.

Several effects of both biological and physical nature contribute to blooms \cite{assmy09}.
From the point of view of biology, phytoplankton control occurs both
top-down (grazing by zooplankton \cite{scheffer97}), and bottom-up (nutrient 
availability; the microbial loop \cite{huppert02}). In many situations, 
both mechanisms are expected to contribute simultaneously to the bloom event
\cite{carbonel99}. 
Physical processes, such as variation of turbidity in the water basin \cite{may03},
modifications in the thermocline \cite{smayda97a}, and mixing by turbulence \cite{huisman99}
may equally contribute to the dynamics.

We shall focus in this paper on the particular route to bloom formation provided
by failure of the top-down control by zooplankton. Central to this is the so called
mismatch issue \cite{cushing90}. since the life cycle of phytoplankton is typically one
order of magnitude shorter than that of zooplankton, a positive fluctuation
in phytoplankton productivity is not compensated by a simultaneous
increase in the zooplankton population. This produces the bloom event:
the phytoplankton population 
escapes zooplankton control and grows to the carrying capacity of the medium,
before also the zooplankton population grows appreciably, and is able to 
bring that of the phytoplankton back to its pre-bloom level.

At a sufficiently coarse grained scale, the phytoplankton-zooplankton 
dynamics can be described by concentration
fields $\bar P(\x,t)$ and $\bar Z(\x,t)$ \cite{evans85,truscott94}. 
This constitutes a mean field approximation
for a stochastic individual dynamics; at different levels of complexity, 
such a description may include the nutrient and even the detritus concentration
fields $\bar N(\x,t)$ and $\bar D(\x,t)$ (NPZ \cite{steele81,fasham90}
and NPZD \cite{edwards01} models). 
The tininess of both phyto- and zooplankton individuals, as well as the size of typical
areas of interest for the study of the concentration dynamics
(at least several meters), suggests that a mean field description 
is indeed the most appropriate. 
The possibility that bloom events be the outcome of 
fluctuations in the system, however, should not be discarded. 
This was the situation observed e.g. in \cite{freund06}, in which 
noise in the external parameters of a compartmental PZ 
model, was able indeed to produce shifts between bloom and no-bloom cycles.

The noise induced shift between limit cycles in a dynamical system
is a well-known effect (see e.g \cite{keeling01} for an application to epidemic spreading).
In most cases the noise considered is external, but recently there has
been a surge of interest in the role of internally generated demographic noise 
(see e.g. \cite{black10} and references therein).
The point that we want to examine in the present paper is precisely whether microscopic
fluctuations, disregarded in a mean field approach, can resurface at macroscopic
scale, generating an internal noise component, sufficient to trigger 
the bloom event. 
This is an example
of demographic fluctuation induced break-up in the mean field description
of reaction-diffusion systems, a type of phenomenon
that have received a great deal of attention in recent years
\cite{young01,mckane05,doering05,butler09,butler11}.
(Notice that, contrary to e.g. \cite{black10}, the noise contribution examined
here acts locally in space).  

The simple PZ model by Truscott and Brindley (TB model), that is going to be considered
here, has the nice characteristic that the plankton behaves like an 
excitable medium \cite{truscott94}. This leads one to expect that noise will be
particularly effective in destabilizing its dynamics. (In fact, \cite{freund06}
considered precisely this model, to generate noise-induced shifts
between seasonal bloom and no-bloom cycles).  In its original version, the TB
model is zero-dimensional, but it can be easily generalized to include 
spatial effects such as advection and diffusion \cite{siekmann08}. 
In the present analysis, a spatially
homogeneous domain will be considered, with no advection, but with
diffusive terms accounting for small scale motions in the water column.
The game to play will be to determine the demographic fluctuations
implicitly neglected in the TB model,
and examine under what conditions they can destabilize the system 
at the level of spatial averages.

This paper is organized as follows. In Sec. II, the main properties of the TB
model are reviewed. In Sec. III, the expression for the stochastic contribution to 
demography are derived. In Sections IV and V, the effect of demographic noise on the
dynamics of the TB model, without and with seasonal forcing, is analyzed. Section VI
is devoted to the conclusions. Some technical details on the master equation treatment of 
demographic fluctuations, in spatially extended domains, are provided  
in the Appendix for reference.

\section{The Truscott-Brindley model}
\label{sec2}
We review here the main properties of the TB model. 
For the moment we disregard the spatial structure of the fields
and focus on the original zero-dimensional version of the model,
which is described by the equations:
\beq
\dot{\bar P}&=&r_0\bar P\Big(1-\frac{\bar P}{K}\Big)-R_m\frac{\bar P^2\bar Z}{\bar P^2+\alpha^2},
\nonumber
\\
\dot{\bar Z}&=&-\mu\bar Z+\gamma R_m\frac{\bar P^2\bar Z}{\bar P^2+\alpha^2}.
\label{TB}
\eeq
The main characteristics of the dynamics are the following:
\begin{itemize}
\item
Logistic reproductive behavior of the phytoplankton,
with the carrying capacity $K$ determining the maximum 
concentration that the medium can support at steady state.
\item
Grazing by zooplankton characterized by a so 
called Holling-III kind of behavior \cite{holling59}. Zooplankton are able to graze
on phytoplankton with optimal rate $R_m\bar Z$, only if the concentration
of the second is above the level fixed by the half-saturation concentration
$\alpha$. Below this threshold, the grazing rate is quadratic in $\bar P$:
$R_m(\bar P/\alpha)^2\bar Z$, reflecting both a reduced grazing ability
of the zooplankton in a dilute environment, and the actual reduced
amount of food available.
\item
A carrying capacity of the medium supposed much larger than the half saturation 
concentration, $K\gg\alpha$, meaning that at high values of $\bar P$, 
the ability of zooplankton to control phytoplankton growth is limited.
\item
A zooplankton reproductive dynamics supposed slower than that of the
phytoplankton: $\mu/R_m\ll 1$, while $r_0/R_m\sim 1$.
A conversion efficiency $\gamma$ assumed consistently
small.
\end{itemize}

\begin{figure}
\begin{center}
\includegraphics[draft=false,width=8cm]{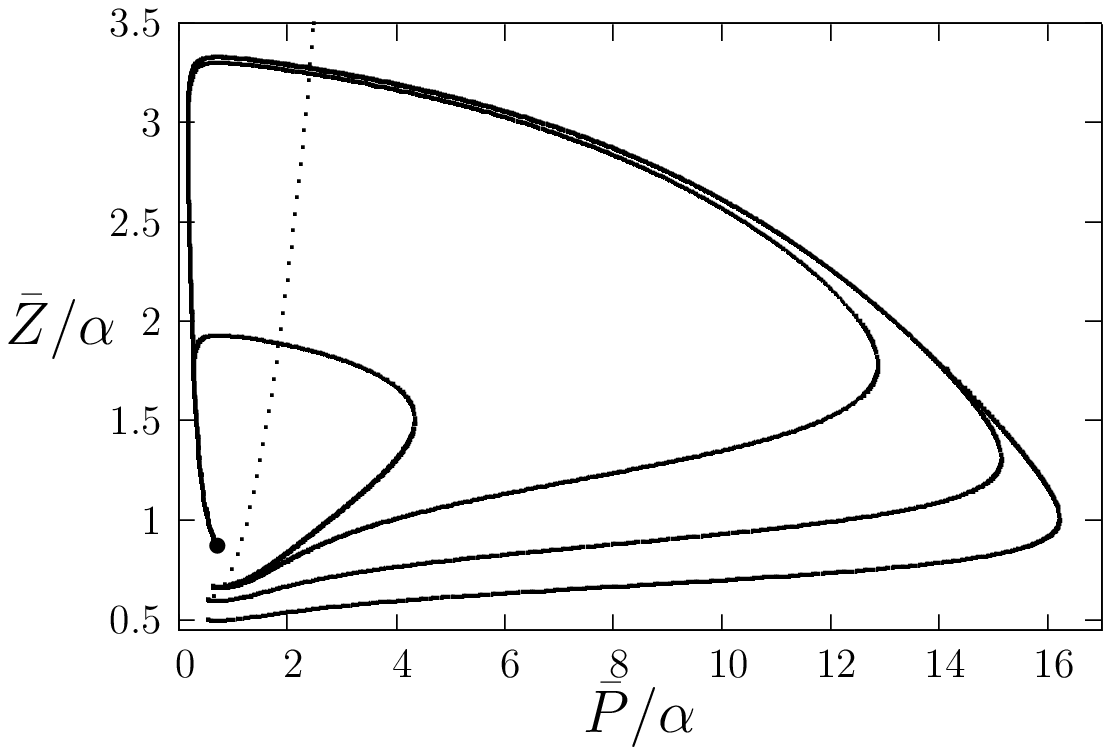}
\caption
{
Trajectories approaching the fixed point $(P_f,Z_f)$ (fat dot to lower left corner of
picture) starting from different initial conditions; going from outer to inner trajectory:
$(\bar P,\bar Z)=(0.5,0.5);\ (0.5.0.6);\ (0.6,0.66);\ (0.6,0.68)$. 
The parameters are those of Eq. (\ref{paramdim}). The dotted line indicate change of
signature of the Jacobian matrix for Eq. (\ref{TB}), and could 
roughly be identified as the point where the bloom begins.
}
\label{trusfig1}
\end{center}
\end{figure}

The default values of the constants that are utilized in the zero-dimensional
case are \cite{truscott94}:
\beq
&&r_0=0.3/{\rm day},\quad
R_m=0.7/{\rm day},\quad
\mu=0.012/{\rm day},
\nonumber
\\
&&K=108{\rm mg\, C/m^3},
\quad
\alpha=5.7{\rm mg\, C/m^3},
\label{paramdim}
\\
&&\gamma=0.05,
\nonumber
\eeq
where the units ``${\rm mg\, C}$'' stand for carbon milligrams in dry weight.
From here we can extract three independent dimensionless groups
\beq
&&\hat r_0=\frac{r_0}{R_m}\simeq 0.43,
\quad
q=\frac{\mu}{\gamma R_m}\simeq 0.34,
\nonumber
\\
&&\epsilon=\frac{\alpha}{K}\simeq 0.053.
\label{param}
\eeq
We see that the dynamics is characterized by
two independent small parameters: $\gamma$ and $\epsilon$.
The dynamics described by Eq. (\ref{TB}) 
has a fixed point at the scale of the half-saturation 
constant $\alpha$:
\beq
P_f&=&\alpha\sqrt{\frac{q}{1-q}},
\nonumber
\\
Z_f&=&\frac{\hat r_0}{P_f}(1-\frac{P_f}{K})(\alpha^2+P_f^2)
\simeq\frac{\alpha\hat r_0}{\sqrt{q(1-q)}}.
\label{fp}
\eeq
For small $\gamma,\epsilon\ll 1$ and $q<1/2$, this fixed point is globally attracting
(a Holling-III functional form for grazing appears to be crucial for stability).
However, if the initial zooplankton concentration is too low,  
before reaching the fixed point, the system
will make an excursion to the high $\bar P\sim K$ range,
which could be interpreted as a bloom event.
The situation is illustrated in Fig. \ref{trusfig1}. 
As shown in figure, the onset of bloom could roughly be
identified in the $\bar P\bar Z$ plane by the line where
the largest eigenvalue of the Jacobian of Eq. (\ref{TB})
crosses to positive, and phase points start to separate exponentially.

The above picture of bloom triggering by zooplankton depletion 
can be improved including the effect of seasonal forcing. Model equations (\ref{TB})
can accommodate this effect by letting the phytoplankton productivity $r$ become
dependent on the temperature, as suggested in \cite{freund06}. The parameterization that we adopt 
is the same as in \cite{freund06}: a Van't Hoff kind of dependence for $r$ \cite{berges02}:
\beq
r_0\to r(T)=r_0\ 2^{\upsilon(T-T_0)}
\label{vanthoft}
\eeq
and a sinusoidal dependence on time of the temperature:
\beq
T(t)=T_0+\Delta T\sin(\Omega t+\phi).
\label{DeltaT}
\eeq
with $\Omega=2\pi/(365\ {\rm days})$ to allow for an annual cycle, 
and $\phi/(2\pi)=0.59$, to have that setting $t=0$ on  January 1st, causes
the first temperature minimum to occur on March 1st and the first maximum 
on August 29th. For the sake of definiteness, as in \cite{freund06}, we set
$\Delta T=6{\rm ^oC}$ and $\upsilon=0.1\,{\rm ^oC}^{-1}$.

\begin{figure}
\begin{center}
\includegraphics[draft=false,width=8cm]{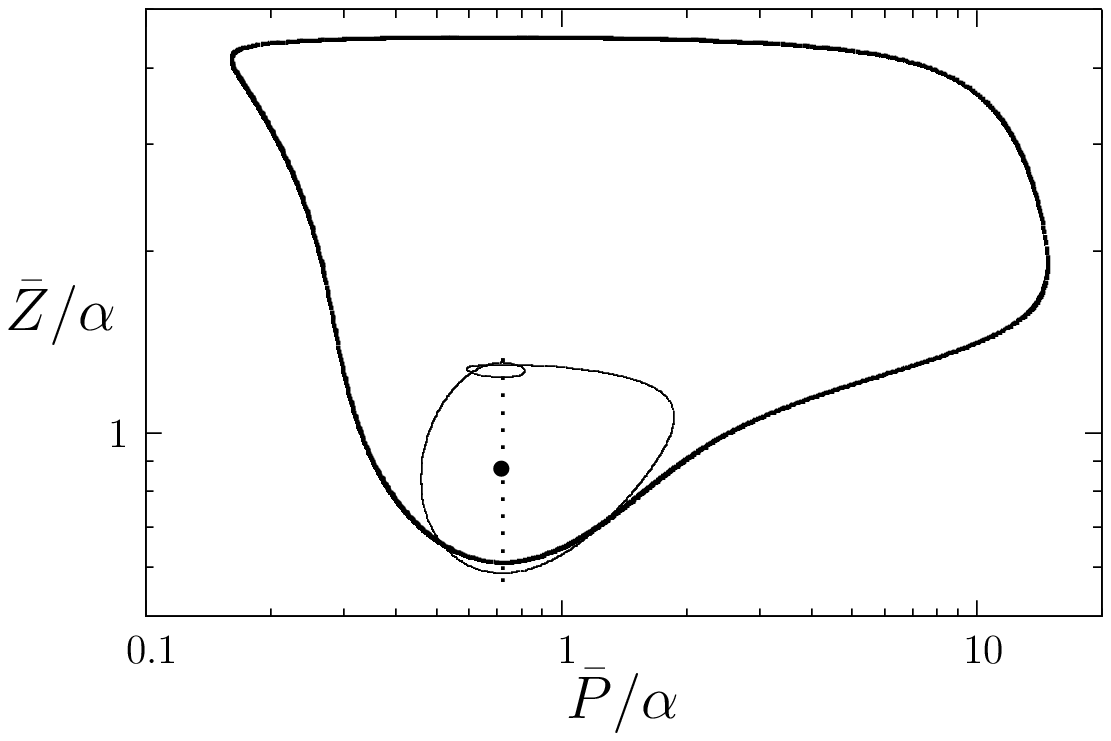}
\caption
{
Bloom and no-bloom cycles (heavy and thin lines respectively) in the seasonally forced
TB model.  The phase points circle counterclockwise along the cycle. 
Notice that the intersections of the two cycles corresponds to phese points that in the
two cycles are associated to different (although close) instants of time.
The dotted vertical line gives the position of the fixed points that would correspond to 
the different values taken by $\hat r$ during the year. 
The fat dot still identifies the fixed point corresponding to  $\hat r=\hat r_0$.
The parameters in the graph are those in Eqs. (\ref{vanthoft}-\ref{DeltaT}). 
}
\label{trusfig2}
\end{center}
\end{figure}

\begin{figure}
\begin{center}
\includegraphics[draft=false,width=8cm]{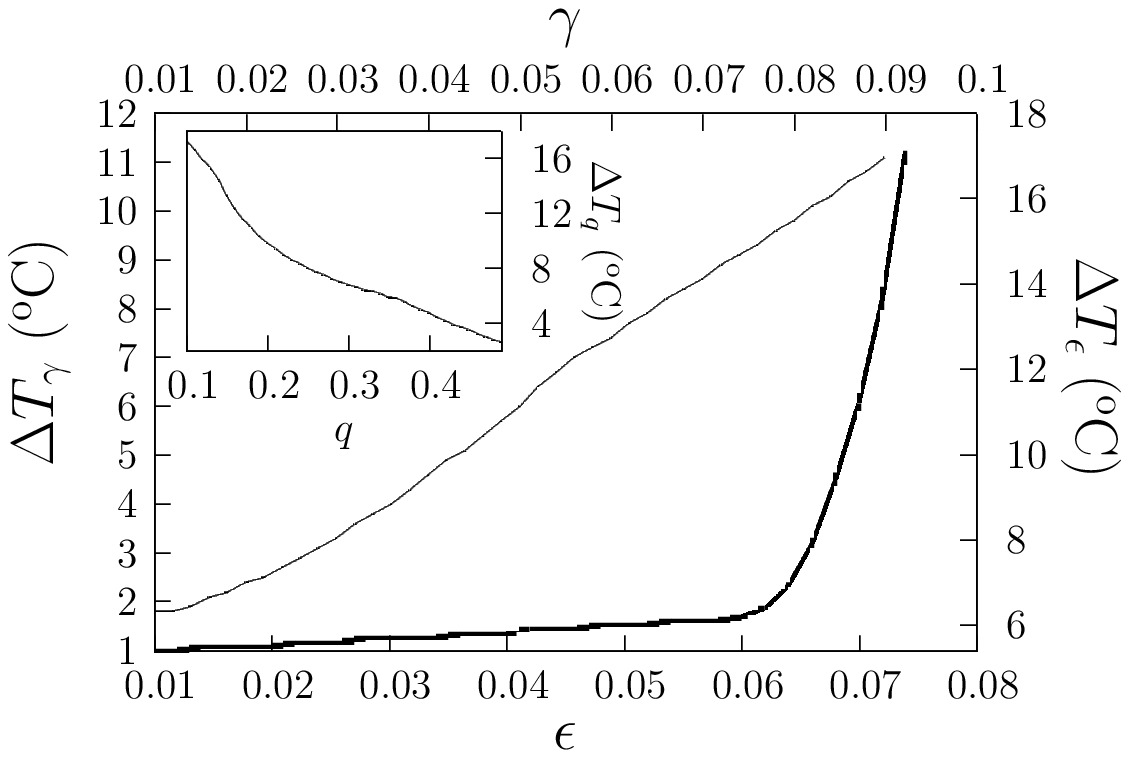}
\caption
{
Dependence on the critical forcing amplitude $\Delta T_{crit}$ on the parameters
$\gamma$ (thin line), $\epsilon$ (heavy line) and $q$ (insert). In the three cases
$\Delta T_{\gamma,\epsilon,q}$ indicate the value of $\Delta T_{crit}$ obtained
varying $\gamma,\epsilon,q$ respectively and keeping the remaining parameters fixed
to the values in Eq. (\ref{param}).
}
\label{trusfig3}
\end{center}
\end{figure}

Adding a seasonal forcing, 
turns out to modify the dynamics in important way, with the single fixed point
in the autonomous case leaving way to two stable limit cycles \cite{freund06}.
The situation is illustrated in Fig. \ref{trusfig2}: a small amplitude no-bloom
cycle coexists with a large amplitude bloom cycle, each one characterized
by a well defined (time-dependent) basin of attraction.
%
%
%
As illustrated in Fig. \ref{trusfig3}, the small cycle will remain stable only if the
seasonal temperature excursion $\Delta T$ is below a critical threshold
$\Delta T_{crit}(\hat r_0,q,\epsilon,\gamma)$, that, for the values of the parameters
quoted in Eqs. (\ref{paramdim}) is $\Delta T_{crit}\simeq 6.1{\rm ^oC}$. 
Similarly, it can be shown that,  if $\Delta T$ is too small, only the no-bloom cycle
will survive.
The no-bloom destabilization 
threshold is very close to the value of the forcing $\Delta T=6{\rm ^oC}$ considered
in \cite{freund06}. The analysis in that paper showed in fact that
addition of a fluctuating component to the forcing, lead to random
switch from year to year between the two regimes.
As it is clear from Fig. \ref{trusfig2}, the destabilization is likely to take
place near the intersection between the bloom and no-bloom trajectories,
where the separation between the phase points of the two cycles (at equal 
times) is smaller.


%
%
%
\section{Demographic fluctuations}
\label{sec3}
We consider a situation in which the typical size of a zooplankton individual
is much larger
than that of a typical phytoplankter. A reasonable estimate (with large variations) 
for the mass of a copepod could be, for instance:
\beq
m_Z\sim 20{\rm \mu g\, C},
\label{m_Z}
\eeq
corresponding to a typical individual size in the millimeter range \cite{gonzalez08,metcalfe04}. 
In a no-bloom regime $\bar P,\bar Z\sim\alpha$ with $\alpha$ given as
in Eq. (\ref{paramdim}), Eq. (\ref{m_Z}) would lead to a numerical density 
of the order of one zooplankton individual per liter. Given the
much smaller size of microscopic algae, phytoplankton could in turn be
treated as a continuum at those scales. We thus expect that the 
demographic fluctuations in the system be driven by the zooplankton.

Locally, demographic fluctuations will be the result of a competition between
stochasticity at the individual level of the birth-death process, and mixing
by spatial transport. 
We shall 
consider a two-dimensional situation, in which the vertical structure
of the water column is not resolved, and parameterize horizontal
mixing through a diffusivity $\kappa$, supposed equal for both phyto- and
zooplankton.
This is probably the only viable strategy to describe a very
complex situation, in which microscopic swimming, stirring by
larger organisms, turbulence generated by perturbations at the water surface,
all play an important role \cite{note}. Likewise, we shall neglect all effects of 
large scale advection, including the forcing induced by the formation of
fronts, where the two plankton populations may get out of balance \cite{reigada03}. 

Including the possibility of a 2D spatial structure, the original model equations
(\ref{TB}) can be written in the form
\beq
\dot{\bar P}&=&B_P(\bar P,\bar Z)-D_P(\bar P,\bar Z)+\kappa\nabla^2\bar P,
\nonumber
\\
\dot{\bar Z}&=&B_Z(\bar P,\bar Z)-D_Z(\bar P,\bar Z)+\kappa\nabla^2\bar Z,
\label{TB1}
\eeq
where $B_{PZ}$ and $D_{PZ}$ give the local birth and death rates in the population,
and $\bar P=\bar P(\x,t)$ and $\bar Z=\bar Z(\x,t)$ give the plankton concentration
averaged over the height of the water column.
Expressing time in units $R_m^{-1}$ and concentrations in units $\alpha$ 
(and lengths in terms of some reference scale), the parameters
entering Eq. (\ref{TB1}) can be written in the form
\beq 
&&B_P=\hat r\bar P;
\qquad
D_P=\epsilon\hat r\bar P^2+\bar P^2\bar Z/(1+\bar P^2);
\nonumber
\\
&&B_Z=\gamma\bar P^2\bar Z/(1+\bar P^2);
\qquad
D_Z=\gamma q\bar Z.
\label{BD}
\eeq
From now on, unless otherwise stated, all relations will be expressed in dimensionless
form.

Rather than working in a field theoretical setting \cite{doi76}, we prefer to apply
the standard system size expansion approach of van Kampen directly to the master equation
for the system \cite{vankampen}.
Let us therefore partition the domain in volumes of horizontal size $\Delta x$,
each one containing instantaneously $N_P,N_Z$ 
individuals of the $P$ and $Z$ groups.  We can introduce instantaneous
concentrations $P,Z=\Omega_{P,Z}^{-1}N_{P,Z}$,
coarse grained at horizontal scale $\Delta x$, with 
\beq
\Omega_{P,Z}=h(\Delta x)^2/m_{P,Z}
\label{Omega}
\eeq
parameterizing the size of the population in the volumes, and $h$ indicating
the height of the water column. 
Indicate with $\tilde P=P-\bar P$ and $\tilde Z=Z-\bar Z$ the fluctuating part of
the coarse-grained field;
in the present situation of fluctuations driven by zooplankton discreteness,
we expect $\tilde P/\bar P,\tilde Z/\bar Z=O(N_Z^{-1/2})$. Let us
define normalized fluctuation fields $\phi,\zeta$ with this scaling contribution
factored out:
\beq
N_P&=&\Omega_P P:=\Omega_P(\bar P+\Omega_Z^{-1/2}\phi),
\nonumber
\\
N_Z&=&\Omega_Z Z:=\Omega_Z(\bar Z+\Omega_Z^{-1/2}\zeta).
\nonumber
\eeq

Following the strategy utilized in \cite{butler09,butler11,bonachela12}, we adopt
as modelling assumption that the birth and death rates $B_{P,Z}$ and $D_{P,Z}$
at the population scale, reflect birth and death rates at the individual level, conditioned
to the instantaneous values of the concentration fields $P(\x,t)$ and $Z(\x,t)$. 
An hypothesis of independence and Markovianity of the birth-death events, at 
scales of interest, underlies this assumption. 
At the population
level, the transition probabilities $\d\mathcal{P}$ in an interval $\d t$ will be
therefore:
\beq
&N_P\to N_P+1:&\d\mathcal{P}=\Omega_PB_P(P,Z)\d t,
\nonumber
\\
&N_P\to N_P-1:&\d\mathcal{P}=\Omega_PD_P(P,Z)\d t,
\nonumber
\\
&N_Z\to N_Z+1:&\d\mathcal{P}=\Omega_ZB_Z(P,Z)\d t,
\nonumber
\\
&N_Z\to N_Z-1:&\d\mathcal{P}=\Omega_ZD_Z(P,Z)\d t,
\label{prob}
\eeq 
and $B_P/P,\ldots D_Z/Z$ will be the corresponding birth and death probabilities per unit
time for the individuals.

A standard procedure \cite{vankampen}
leads, from Eq. (\ref{prob}), to the master equation for
the probability density function (PDF) $\rho(\{\phi_i,\zeta_i\},t)$ 
(the index $i$ labels the volumes in the domain):
\beq
\frac{\partial\rho}{\partial t}
&=&\sum_i\left\{\Pi_i\rho+
\Omega_Z^{1/2}\Big[\dot{\bar P}_i\frac{\partial\rho}{\partial\phi_i}+
\dot{\bar Z}_i\frac{\partial\rho}{\partial\zeta_i}\Big]\right.
\nonumber
\\
&+&
\Omega_P\Big\{\Big[\exp\Big(-\Omega_C^{-1/2}\frac{\partial}{\partial\phi_i}\Big)-1\Big]B_{P,i}
\nonumber
\\
&+&\Big[\exp\Big(\Omega_C^{-1/2}\frac{\partial}{\partial\phi_i}\Big)-1\Big]D_{P,i}\Big\}\rho
\nonumber
\\
&+&
\Omega_Z\Big\{\Big[\exp\Big(-\Omega_Z^{-1/2}\frac{\partial}{\partial\zeta_i}\Big)-1\Big]B_{Z,i}
\nonumber
\\
&+&\left.\Big[\exp\Big(\Omega_Z^{-1/2}\frac{\partial}{\partial\zeta_i}\Big)-1\Big]
D_{Z,i}\Big\}\rho
\right\},
\label{master}
\eeq
where $\Omega_C^{1/2}=\Omega_Z^{-1/2}\Omega_P$, and the
additional term $\Pi_i\rho$ accounts for the exchange of plankton between
adjacent volumes produced by diffusion (see Appendix). 
Expanding to lowest order in $\Omega_{P,Z}^{-1}$, both the exponentials
and the reaction rates 
in Eq. (\ref{master}) [recall that 
$B_P=B_P(\bar P+\Omega^{-1/2}_Z\phi,\bar Z+\Omega^{-1/2}_Z\zeta),\ldots$],
we obtain the Fokker-Planck equation:
\beq
\frac{\partial\rho}{\partial t}&+&
\sum_i\Big\{
\frac{\partial}{\partial\phi_i}\Big(\aPPi\phi_i+\aPZi\zeta_i\Big)\rho
\nonumber
\\
&+&\frac{\partial}{\partial\zeta_i}\Big(\aZPi\phi_i+\aZZi\zeta_i\Big)\rho\Big\}
\nonumber
\\
&=&
\sum_i\Pi_i\rho+ \frac{1}{2} \sum_{ij}
\frac{\partial}{\partial\zeta_i}\frac{\partial}{\partial\zeta_j}\Xi_{ij}\rho,
\label{Fokker-Planck}
\eeq
where $\Xi_{ij}=
[B_Z(\bar P_i,\bar Z_i)+D_Z(\bar P_i,\bar Z_i)]\delta_{ij}$,
and the $\aPP,\ldots$
give the entries of the Jacobian matrix for Eq. (\ref{TB})
($\aPP\equiv\partial\dot{\bar P}/\partial\bar P,\ldots$). Notice that the
only noise correlator present in the equation, $\Xi_{ij}$, is the one
associated with the variable $\zeta$, i.e. with the zooplankton fluctuations.

The dependence of the expansion parameter $\Omega_Z^{-1}$ on the
arbitrary coarse graining scale $\Delta x$ may look somewhat confusing, 
as different $\Delta x$ correspond to different
fluctuation levels at the coarse-graining scale. We see
in particular that $\Omega_Z$ goes to zero with $\Delta x$. 
Nevertheless, as it will become
clear in the next section, there exists a natural microscopic scale $\lambda_c$ below
which fluctuations are smeared out by diffusion. Thus, applicability
of the system size expansion rests on smallness of fluctuations at scale $\lambda_c$,
and not on the choice of $\Delta x$.

Going back to the original variables $\tilde P,\tilde Z$, 
and taking the continuous limit, we see that the Fokker-Planck equation (\ref{Fokker-Planck})
is equivalent to the system of Langevin equations
\beq 
&&\frac{\partial\tilde P}{\partial t}+\aPP\tilde P+\aPZ\tilde Z=
\kappa\nabla^2\tilde P
\nonumber
\\
&&\frac{\partial\tilde Z}{\partial t}+\aZP\tilde P+\aZZ\tilde Z=\kappa\nabla^2\tilde Z+\xi,
\label{Langevin}
\eeq
where 
\beq
&&\langle\xi(\x,t)\xi(\x',0)\rangle=\Xi(\x,t)\delta(\x-\x')\delta(t),
\nonumber
\\
&&\Xi(\x,t)=\hat m_Z[B_Z(\bar P,\bar Z)+D_Z(\bar P,\bar Z)],
\label{noise}
\eeq
and we have put $\hat m_Z=m_Z/h$, 
$\bar P\equiv\bar P(\x,t)$ and $\bar Z\equiv\bar Z(\x,t)$ (more details in the Appendix).

%
%
%
%
\section{Dynamics near the fixed point}
\label{sec4}
Let us consider first the case of a TB model without seasonal forcing, and
study the fluctuations around the fixed point given by Eq. (\ref{fp}). The analysis is
similar to the one carried on in \cite{butler11} on another PZ model (the Levin-Segel model 
\cite{levin76}), which focused on the destabilization of Turing patterns (see also 
\cite{biancalani11} for another example of Turing pattern destabilization
a reaction-difusion system). The calculation
will allow to identify the relevant fluctuation scales on which 
to base the forced case analysis in the next section.

The evolution equations
for the correlation functions $C_{PP}(\x,t)=\langle\tilde P(\x,t)\tilde P(0,t)\rangle,\ldots$
can be obtained from Eqs. (\ref{Langevin}-\ref{noise}), and take the form in Fourier space:
\beq
&&\frac{1}{2}\dot C_{PP\k}+(\aPP+\kappa k^2)C_{PP\k}+\aPZ C_{PZ\k}=0
\nonumber
\\
&&\dot C_{PZ\k}+\aZP C_{PP\k}+(\aPP+\aZZ+2\kappa k^2)C_{PZ\k}
\nonumber
\\
&&\qquad\;\;\; +\aPZ C_{ZZ\k}=0
\nonumber
\\
&&\frac{1}{2}\dot C_{ZZ\k}+\aZP C_{PZ\k}+(\aZZ+\kappa k^2)C_{ZZ\k}
=\Xi.
\label{corr}
\eeq
At the fixed point we find immediately, using Eq. (\ref{BD}) and 
working to lowest order in $\epsilon$ and $\gamma$:
\beq
&&\aPP\simeq -\hat r(1-2q),
\quad
\aPZ=-q,
\nonumber
\\
&&\aZP\simeq 2\gamma\hat r(1-q),
\quad
\aZZ=0,
\nonumber
\\
&&
\Xi\simeq 2\gamma\hat m_Z\sqrt{q/(1-q)}.
\label{param1}
\eeq
Equation (\ref{corr}) becomes at steady state:
\beq
&&(\aPP+\kappa k^2)C_{PP\k}+\aPZ C_{PZ\k}=0,
\nonumber
\\
&&\aZP C_{PP\k}+(\aPP+2\kappa k^2)C_{PZ\k}+\aPZ C_{ZZ\k}=0,
\nonumber
\\
&&\aZP C_{PZ\k}+\kappa k^2C_{ZZ\k}=\Xi,
\nonumber
\eeq
that has solution
\beq
C_{PP\k}&\simeq&\Delta^{-1}\aPZ^2\Xi,
\nonumber
\\
C_{PZ\k}&\simeq&\Delta^{-1}\aPZ(-\aPP+\kappa k^2)\Xi,
\nonumber
\\
C_{ZZ\k}&\simeq&\Delta^{-1}(-\aPP+\kappa k^2)^2\Xi,
\label{sol}
\eeq
with 
\beq
\Delta=\aPP\aPZ\aZP+\aPP^2\kappa k^2-2\aPP\kappa^2k^4+\kappa^3k^6.
\label{discr}
\eeq
From Eqs. (\ref{sol}-\ref{discr}), we see that there is
a long wavelength range, dominated by
demography; using Eq. (\ref{param1}):
\beq
C_{PP\k}&\simeq&\frac{q\hat m_Z}{\hat r_0^2(1-q)(1-2q)}\sqrt{\frac{q}{1-q}},
\nonumber
\\
C_{PZ\k}&\simeq&\frac{\hat m_Z}{\hat r_0(1-q)}\sqrt{\frac{q}{1-q}},
\nonumber
\\
C_{ZZ\k}&\simeq&\frac{\hat m_Z(1-2q)}{q(1-q)}\sqrt{\frac{q}{1-q}}.
\label{large_scale}
\eeq
At small scales, the fluctuations are smeared out by diffusion, 
with the asymptotic
law $C_{PP\k}\simeq\aPP^2\Xi/(\kappa k^2)^3$, $C_{PZ\k}\simeq\aPZ\Xi/(\kappa k^2)^2$,
$C_{PZ\k}\simeq\Xi/(\kappa k^2)$. The transition occurs at
$\kappa k^2\sim\aZP\aPZ/\aPP$, which sets the crossover length, from 
Eqs. (\ref{param}) and (\ref{param1}), back to dimensional units:
\beq
\lambda_c=\sqrt{\kappa/\mu}.
\label{lambda_c}
\eeq
This is the typical distance travelled by
a zooplankter in a lifetime and corresponds to the characteristic wavelength of the 
chemical waves supported by the system in the mean field. 
Notice that, contrary to the case considered in \cite{siekmann08}, for the choice of parameters
utilized, no Turing instability is present.

The correlation spectrum that has been obtained, characterized by a plateau at $k\lambda_c<1$,
and a decay at $k\lambda_c>1$, corresponds to fluctuations with a correlation scale
$\lambda_c$. The fluctuation amplitude can be estimated approximating the decay at 
$k\lambda_c>1$ with a step function and approximating the solution for $k\lambda_c<1$
with Eq. (\ref{large_scale}). This gives for the fluctuation amplitude
\beq
C_{PP}(0)=\mathcal{F}^{-1}_{\x=0}[C_{PP\k}]\sim\lambda_c^{-2}C_{PP0},
\nonumber
\eeq 
with $C_{PP0}$ as given in Eq. (\ref{large_scale}),
and similar expressions for $C_{PZ}(0)$ and $C_{ZZ}(0)$. From Eqs. (\ref{large_scale})
we get for the ratio of the fluctuation amplitude to the mean (back to dimensional units):
\beq
\frac{C_{PP}(0)}{\bar P^2}\sim\frac{C_{PZ}(0)}{\bar P\bar Z}
\sim \frac{C_{ZZ}(0)}{\bar Z^2}\sim\frac{m_Z}{\alpha h\lambda_c^2},
\label{estimate}
\eeq
that is the ratio of the zooplankter mass and the typical total zooplankton 
mass in a water column of height $h$ and horizontal extension $\lambda_c$.

\section{Destabilization of the no-bloom regime}
\label{sec5}
Let us pass to consider a seasonally forced situation and ask under what conditions,
demographic noise could destabilize global bloom and no-bloom cycles.

Clearly, the linearized theory of Sec. \ref{sec4} cannot be utilized 
in the present case, as the trajectories evolve for most of their time in an unstable region,
as depicted in Fig. \ref{trusfig1}. A first possibility is at this point
numerical solution of the master equation (\ref{master}) by Montecarlo techniques,
utilizing the transition probabilities in  Eq. (\ref{prob}) to deal with the birth and
death events in the computational cells $\Delta x$, and the transfer rates $W_i$ introduced
in Appendix A to deal with diffusion. In alternative, 
the Langevin equation (\ref{Langevin}) can be integrated numerically, after
replacement of the Jacobian matrix $a_{ij}$ with the full RHS (right hand side) of 
Eq. (\ref{TB1}), to extend beyond linear regime the region of validity of 
the equation

It is convenient to express lengths
in units $\lambda_c$, so that the forced equation can be written in the form
\beq
\dot{P}&=&B_P(P,Z)-D_P(P,Z)+\gamma q\nabla^2P,
\nonumber
\\
\dot{Z}&=&B_Z(P,Z)-D_Z(P,Z)+\gamma q\nabla^2Z+\xi.
\label{TB2}
\eeq
Notice the dependence of the terms on RHS of Eq. (\ref{TB2}) on the fluctuating fields 
$P$ and $Z$; similarly, the dependence on $\bar P$ and $\bar Z$ in Eq. (\ref{noise}) 
is replaced by one on $P$ and $Z$:
\beq
&&\langle\xi(\x,t)\xi(\x',0)\rangle=\Xi(\x,t)\delta(\x-\x')\delta(t),
\nonumber
\\
&&\Xi(\x,t)=\hat m_Z[B_Z(P,Z)+D_Z(P,Z)].
\label{noise1}
\eeq
The parameter $\hat m_Z$, which, through Eq. (\ref{noise1}), 
determines the amplitude of the noise $\xi$, takes the form, in terms of dimensional parameters:
\beq
\hat m_Z=\frac{m_Z\mu}{\alpha h\kappa}.
\label{hat m_Z}
\eeq
This quantity will play the role of a control parameter for the theory.
[We notice by the way that 
$\hat m_Z$ coincides with the amplitude ratio in Eq. (\ref{estimate})].

The results that follow come from direct numerical simulation of Eqs. (\ref{TB2}-\ref{noise1})
in a periodic domain
using a simple finite difference scheme (centered in space, forward Euler in time).
Except for very large values of $\hat m_Z$, the results can be shown to coincide with those 
of Montecarlo simulation \cite{montecarlo}.
(For larger values of $\hat m_Z$,
the discretization $\Delta x<\lambda_c$ would become small enough for local 
extinction in the computation cells $\Delta x$ to become a problem, and either Montecarlo,
or more sophisticated algorithms, such as those described in \cite{dornic05,moro04},
should be utilized).

\begin{figure}
\begin{center}
\includegraphics[draft=false,width=7cm]{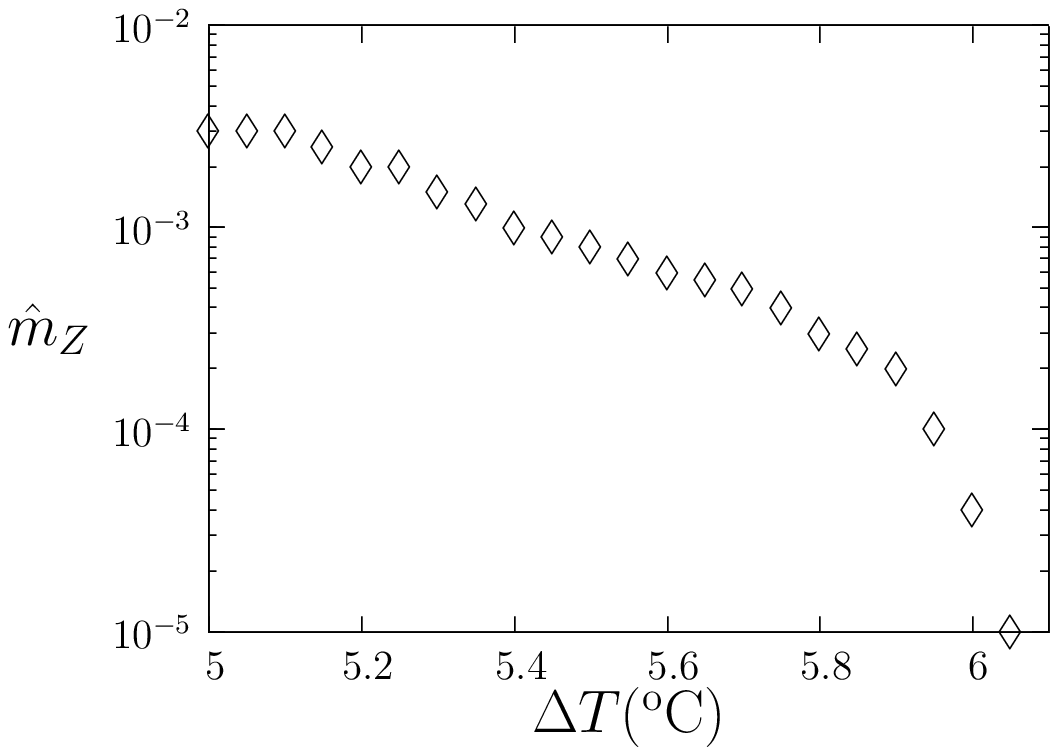}
\caption
{
No-bloom cycle destabilization threshold in function of $\Delta T$, from numerical
simulation in a periodic domain $200\Delta x\times 200\Delta x$ with $\Delta x=\lambda_c/10$.
The no-bloom cycle is considered destabilized if the system crosses the threshold $\bar P=10$
before $t=10\,$years. All parameters except $\Delta T$ and $\hat m_Z$ set
to the values in Eqs. (\ref{paramdim}) and (\ref{m_Z}). Initial conditions set equal to
$(P_f,Z_f)$ uniformly in the domain.
}
\label{trusfig4}
\end{center}
\end{figure}

As expected, a sufficiently high level of noise destabilizes the small cycle and leads
to locking the system in the large bloom cycle. As illustrated in Fig. \ref{trusfig4}, the
threshold in $\hat m_Z$ becomes lower as the critical forcing $\Delta T_{crit}$ is 
approached. 
\begin{figure}
\begin{center}
\includegraphics[draft=false,width=8.5cm]{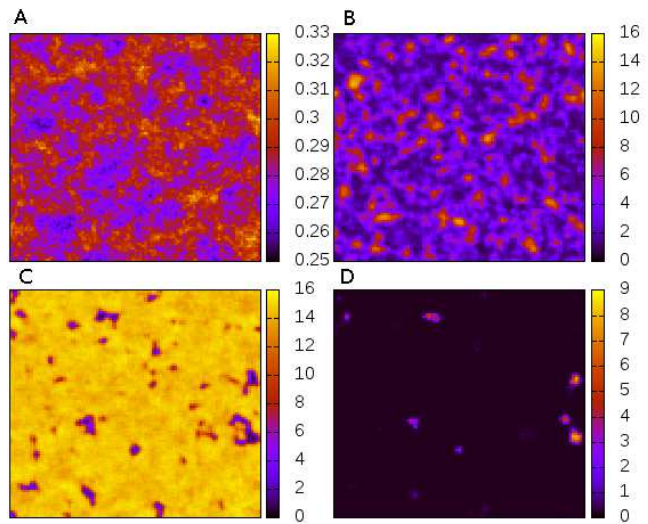}
\caption
{(Color online).
Snapshots of the phytoplankton concentration field from a simulation with $\hat m_Z=10^{-4}$;
$\Delta T=6\,{\rm ^oC}$ and other parameters set as in Eqs. (\ref{paramdim}) and (\ref{m_Z}).
Size of the domain $200\Delta x\times 200\Delta x$; periodic boundary conditions, with 
$\Delta x=0.3\lambda_c$.
Initial conditions set equal to $(P_f,Z_f)$ uniformly in the domain.
A: day 700 (December 1st: no-bloom condition); B: day 890 (May 9th: pre-bloom condition);
C: day 924 (June 13th: bloom peak); D: day 970 (July 29: concentration minimum after bloom).
}
\label{trusfig5}
\end{center}
\end{figure}
\begin{figure}
\begin{center}
\includegraphics[draft=false,width=8.5cm]{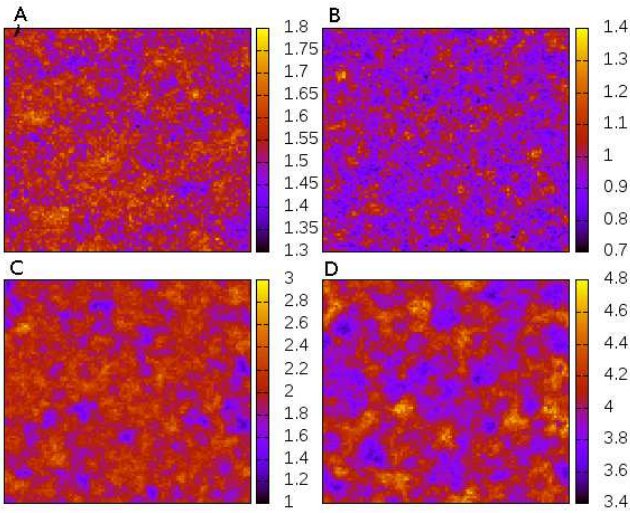}
\caption
{
(Color online).
Same as Fig. \ref{trusfig5} in the case of the zooplankton field.
}
\label{trusfig6}
\end{center}
\end{figure}
\begin{figure}
\begin{center}
\includegraphics[draft=false,width=8.5cm]{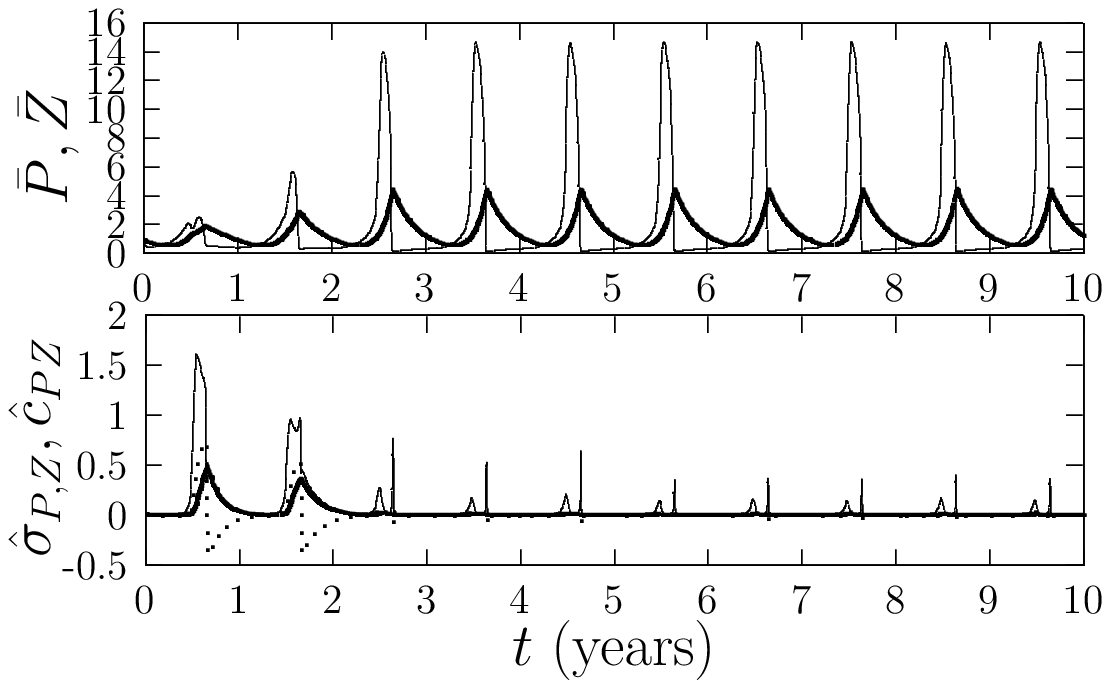}
\caption
{
Top figure: 
evolution of the mean concentration fields $\bar P$ (thin line) and $\bar Z$ (heavy line).
Bottom figure:
evolution of the normalized
RMS fluctuations $\hat\sigma_P=\langle\tilde P^2\rangle^{1/2}/\bar P$ (thin line),
$\hat\sigma_Z=\langle\tilde Z^2\rangle^{1/2}/\bar Z$ (heavy line),
$\hat c_{PZ}=
{\rm sign}(\langle\tilde P\tilde Z\rangle)|\langle\tilde P\tilde Z\rangle/(\bar P\bar Z)|^{1/2}$
(dotted line).
Same choice of parameters as in Figs. \ref{trusfig5} and \ref{trusfig6}.
}
\label{trusfig7}
\end{center}
\end{figure}
\begin{figure}
\begin{center}
\includegraphics[draft=false,width=8.5cm]{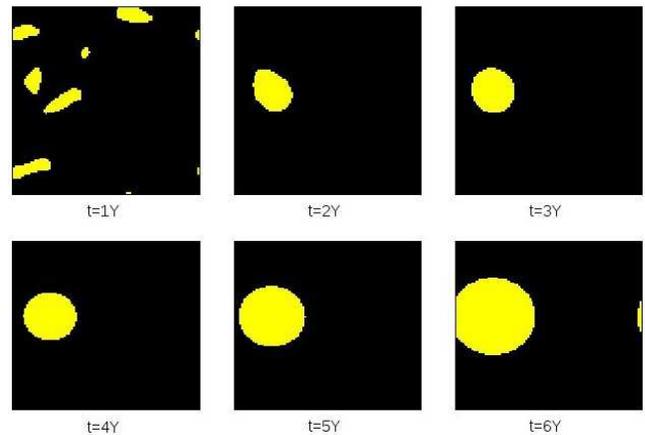}
\caption
{(Color online).
Evolution due to the effect of diffusion, in the absence of noise, 
of domains characterized by bloom dynamics
(light gray in figure -- yellow on web; 
a bloom event at a given pixel is identified by crossing during
the year of the threshold $P=12$). The initial condition at $t=0$ was
a random distribution in space of values $P,Z$ in the bloom and no-bloom
basin of attraction. Initial fraction of bloom points: 0.235 (for larger fractions, the system
locks immediately on a bloom-cycle; the opposite for lower fractions). Domain characteristics
and parameters as in Figs. \ref{trusfig5}-\ref{trusfig6}.
}
\label{trusfig8}
\end{center}
\end{figure}

For default values of the parameters, such as those 
in Eqs. (\ref{paramdim}) and (\ref{m_Z}),  with $\Delta T=6{\rm ^oC}$, 
the threshold would be at $\hat m_Z\simeq 4\cdot 10^{-5}$, which,
for a depth $h\simeq 5\,{\rm m}$, would correspond to a diffusivity
$\kappa\simeq 0.2\,{\rm m^2/day}\equiv 0.024 {\rm cm^2/s}$ and
a correlation length $\lambda_c\simeq 4.1{\rm m}$. In comparison,
$\kappa\sim 0.01\,{\rm cm^2/s}$ would be the diffusivity that would
be produced by microscopic swimming at speed
$\sim 1\,{\rm mm/s}$ with a
persistence time between change of directions of the order 
of one second \cite{metcalfe04}.

As illustrated in Fig. \ref{trusfig5}, even when the system is locked globally
on a bloom cycle, its dynamics is characterized by spatial fluctuations, with regions
of size $\sim\lambda_c$ in which, during bloom events, $P$ remains well below its typical
bloom value.

This picture is confirmed by looking at the evolution of the RMS fluctuations, as depicted 
in Fig. \ref{trusfig7}.
The stronger fluctuation level in the phytoplankton concentration field, at the onset of 
the bloom events and soon after their disappearance, that can be seen in case $B$ and $D$
in Fig. \ref{trusfig5}, is paralleled by the double peaks in $\hat\sigma_P$ in Fig. 
\ref{trusfig7}.

The above picture of global destabilization of the no-bloom cycle is based on numerical
evidence from simulations in a finite domain. One may question whether destabilization
could be just a finite size effect, and would disappear in the thermodynamic limit. One suggestion 
that this is not the case comes from the invasive character of the destabilization 
phenomenon.  The sequence $B-C$ in Fig. \ref{trusfig5} gives a hint of the process:
regions of size $\sim\lambda_c$, characterized by high values
of $P$, through diffusive coupling, 
destabilize nearby regions with low values of $P$, and push them into the bloom cycle.
This picture is corroborated by the behavior of the system in the noiseless $\hat m_Z=0$ case.
As shown in Fig. \ref{trusfig8}, in the absence of noise, the phenomenon persists:
phytoplankton transfer across a $\sim\lambda_c$ distance, from a bloom to a no-bloom region,
destabilizes the second one. Thus, if a bloom bubble is sufficiently large (on the scale
of $\lambda_c$), it will survive diffusive transfer and gradually expand at the expenses
of the surroundings. It is possible to see that 
the situation is confirmed in the case of a system with just two
homogeneous compartmens coupled diffusively, one evolving on a bloom cycle, the other
on a no-bloom cycle: the no-bloom cycle will always be destabilized and the system
will lock on a global bloom cycle.

This suggests the following picture:
\begin{itemize}
\item
Demographic noise continuously generates local fluctuations in $P$ and $Z$ (especially in $Z$)
that may push some regions of space in a bloom regime.
\item
If the noise level is sufficiently high, some of these regions will be large enough for
not being destroyed at once by diffusivity (see passage from year 1 to year 2 in Fig. 
\ref{trusfig8}).
\item
At this point diffusion,
through phytoplankton transfer,
quickly destabilizes the surrounding no-bloom regions.
Noise is not expected to be important any more
in this phase.
\end{itemize}

\section{Conclusions}
The TB model, both in its zero-dimensional version, and in the extended one describing the
evolution of $P$ and $Z$ fields, can be seen as a mean field description of some individual 
level model (ILM).
The most natural way to obtain an individual dynamics from a global
one, is to assume that the birth and death rates at the individual level have the same form
as the corresponding quantities at the population level. Within these assumptions,
independence of the birth and death events, together with absence of memory effects at
the scales of interest, fix the form of the ILM.
A master equation describing its dynamics can thus be obtained, utilizing
techniques analogous to those in \cite{butler09,butler11}. 

It should be stressed that this is only one of the ways in which an ILM could be obtained
from a population level model. Any choice, differing from the present one by identical 
contributions in the individual birth and death rates, 
would work as well, as the equations at the population level would remain unchanged.
The issue is particularly important for the zooplankton, that is the source of 
fluctuations for the model. Nevertheless, the zooplankton death rate $\mu$
appearing the population level equation (\ref{TB}) already reflect modelling 
assumptions at the individual level, so that the present choice of ILM is somewhat 
imposed.

Keeping in mind all these caveats, analysis of the results from the ILM shows that, 
for shallow water conditions (few meters depth), and mixing in the
water column produced by diffusion (diffusivity in the range $0.1\,{\rm m^2/day}$),
demographic noise is sufficient to cause switching between regimes. The effect
is of the same order of magnitude as that of the global temperature
fluctuations considered in \cite{freund06} (fluctuation amplitude $\sim 1\,{\rm ^oC}$
with correlation time equal to 35 days). From the point of view of the PZ model,
this corresponds to a decrease of the instability threshold in the seasonal temperature
forcing, with respect to the mean-field case, of the order of  $1\,{\rm ^oC}$
(see Fig. \ref{trusfig4}). 

An interesting aspect of the present analysis, common to what was obtained in \cite{butler11} in 
the case of Turing waves, is how local fluctuations are able
to produce a global destabilization in the system, that is expected to be maintained
in the thermodynamic limit of an infinite basin.  The reason is 
partly trivial: the reproduction rates at the population level are nonlinear functions
of the concentration fields $P$ and $Z$, and fluctuations lead necessarily to  their
renormalization. Such a picture, however, is incomplete, as the destabilization process,
in the present case,
unless the demographic noise is very large, destabilizes the system only locally. 
(The characteristic scale of the fluctuations coincides with that 
of the Turing patterns of the system, 
and has nothing to do with stochastic demography). 
Only later, by an invasive process, which appears to be insensitive to system size,
the destabilized regions rapidly inglobate the inactive
surroundings, leading to the global bloom cycle. 

It is interesting to notice that, as illustrated in Fig. \ref{trusfig5},
the invasive dynamics is associated with a spatially intermittency 
of the $P$ field in the pre-bloom phase, with the high $P$ regions acting as seeds 
for the coming bloom phase. This condition could be utilized in experiments 
to contrast the particular
destabilization mechanism described here, with the one provided, say,
by global temperature fluctuations.

Again as regards the spatial structure of the blooms,
it is worth comparing the role of the characteristic length $\lambda_c$
defined in Eq. (\ref{lambda_c}), 
with the vertical inhomogeneity scales considered in 
\cite{huisman99}. In both cases, 
the bloom event is produced by a local phytoplankton growth that is not
balanced by dispersion. In one case, horizontally, in the other, in the vertical direction.
The present model bypassed all difficulties associated
with the vertical plankton distribution and vertical structure of the water basin, 
considering implicitly a shallow water condition. An interesting 
question is therefore, whether phenomena such as the ``critical turbulence'' 
in \cite{huisman99} may act as a localization mechanisms, that lets the 
destabilization route described in this paper act also in deep water. 

A different question concerns the robustness of the present results under extension 
of the model to inclusion of bottom-up effects by nutrients from one side,
or inclusion of finer grained details in the PZ dynamics. While one would expect
irrelevance of the $Z$ fluctuations in a 
nutrient unbalance dominated  bloom scenario,
such as the one depicted in \cite{huppert02}, 
the question remains
open as regards the situation in which zooplankton control is dominant.
The feedback by an increase of the zooplankton deaths, on the nutrient field,
and then on the phytoplankton growth rate [the parameter $r_0$ in Eq. (\ref{TB})], would suggest
a positive effect. Closer scrutiny is clearly required.

%


\acknowledgements I wish to thank M. Gatto, R. Casagrandi, A. Lugli\`e and B. Padedda
for interesting and helpful 
discussion. This research was funded in part by Regione Autonoma della Sardegna.


%
%
%
%
%
\appendix
\section{Master equation treatment of diffusion}
We consider for simplicity a one-dimensional domain and a single species, say $Z$. Discretize
the domain and indicate with $N_i$ the number of individuals in slot $i$. We can define the
PDF's:
\beq
\rho_\N(\N)&\equiv&\rho_\N(\{\Omega\bar Z_i+\Omega^{1/2}\zeta_i,i=1,\ldots K\})
\nonumber
\\
&=&\Omega^{-K/2}\rho_\bzeta(\bzeta).
\nonumber
\eeq
Suppose that individuals are transferred diffusively from slot $i$ to slot $i\pm 1$ with a rate
$W_{i\to i+1}=W_{i\to i-1}\equiv W_i=N_i\hat\kappa_i$, where
$\hat\kappa_i=\kappa_i/(\Delta x)^2$, with $\kappa_i\equiv\kappa(\x_i)$ the diffusivity and
$\Delta x$ the width of the slot.
The master equation for $\rho_\N$ can be
written in the form
\beq
\frac{\partial\rho_\N}{\partial t}&=&\sum_i\Big\{
\sum_{k=\pm 1}\exp\Big\{\Omega^{-1/2}\Big(\frac{\partial}{\partial\zeta_{i+k}}
-\frac{\partial}{\partial\zeta_i}\Big)\Big\}
\nonumber
\\
&\times&W_{i+k}
-2W_i\Big\}\rho_\N.
\label{A1}
\eeq
To derive an equation for $\rho\equiv\rho_\bzeta$, 
we exploit the relation
\beq
\Omega^{K/2}\frac{\partial\rho_\N}{\partial t}=
\frac{\partial\rho}{\partial t}
-\Omega^{1/2}\sum_i
\dot{\bar Z}_i\frac{\partial\rho}{\partial\zeta_i}.
\nonumber
\eeq
Substituting into Eq. (\ref{A1}) and expanding  to $O(\Omega^{-1})$,
we obtain:
\beq
\frac{\partial\rho}{\partial t}&=&
\Omega^{1/2}\sum_i
\dot{\bar Z}_i\frac{\partial\rho}{\partial\zeta_i}
\nonumber
\\
&+&\Omega\sum_i\Big\{\Big\{\Big[
1+\frac{1}{\Omega^{1/2}}\Big(\frac{\partial}{\partial\zeta_{i- 1}}
-\frac{\partial}{\partial\zeta_i}\Big)
\nonumber
\\
&+&\frac{1}{2\Omega}\Big(\frac{\partial}{\partial\zeta_{i- 1}}
-\frac{\partial}{\partial\zeta_i}\Big)^2\Big]
w_{i-1}-w_i\Big\}
\nonumber
\\
&+&\Big\{\Big[1+\frac{1}{\Omega^{1/2}}
\Big(\frac{\partial}{\partial\zeta_{i+1}}
-\frac{\partial}{\partial\zeta_i}\Big)
+\frac{1}{2\Omega}
\nonumber
\\
&\times&\Big(\frac{\partial}{\partial\zeta_{i+1}}
-\frac{\partial}{\partial\zeta_i}\Big)^2\Big]
w_{i+1}-w_i\Big\}\Big\}\rho,
\label{A3}
\eeq
where we have introduced the rate density 
$w_i=\Omega^{-1}W_i=Z_i\hat\kappa_i$.
Equation (\ref{A3}) could be further simplified exploiting the
relation 
\beq
\sum_i[w_{i+1}+w_{i-1}-2w_i]\rho=0.
\nonumber
\eeq
At this point we write $w_i=\bar w_i+\Omega^{-1/2}\hat\kappa_i\zeta_i$,
$\bar w_i\equiv\hat\kappa_i\bar Z_i$, and
expand Eq. (\ref{A3}) in powers of $\Omega$. We find, to $O(\Omega^{-1/2})$:
\beq
[\dot{\bar Z}_i-(\bar w_{i+1}+\bar w_{i-1}-2\bar w_i)]
\frac{\partial\rho}{\partial\zeta_i}=0,
\eeq
which gives, taking the continuous limit $\Delta x\to 0$, the diffusion equation
\beq
\frac{\partial\bar Z(x,t)}{\partial t}=\frac{\partial^2(\kappa(x)\bar Z(x,t))}{\partial x^2}.
\label{A6}
\eeq
Going to next order, we find the equation for the fluctuations
\beq
\frac{\partial\rho}{\partial t}&=&\sum_i\Pi_i\rho,
\label{A7}
\eeq
where
\beq
\Pi_i
&=&
\hat\kappa_{i-1}\frac{\partial}{\partial\zeta_{i-1}}\zeta_{i-1}
+\hat\kappa_{i+1}\frac{\partial}{\partial\zeta_{i+1}}\zeta_{i+1}
\nonumber
\\
&-&(\hat\kappa_{i-1}\zeta_{i-1}+\hat\kappa_{i+1}\zeta_{i+1})\frac{\partial}{\partial\zeta_i}
\nonumber
\\
&+&\frac{1}{2}\Big[\hat\kappa_{i-1}\bar Z_{i-1}\Big(\frac{\partial}{\partial\zeta_{i-1}}-
\frac{\partial}{\partial\zeta_i}\Big)^2
\nonumber
\\
&+&\hat\kappa_{i+1}\bar Z_{i+1}\Big(\frac{\partial}{\partial\zeta_{i+1}}-
\frac{\partial}{\partial\zeta_i}\Big)^2\Big].
\label{A8}
\eeq
From Eqs. (\ref{A7}-\ref{A8}) we obtain the equation for the correlation 
$C_{ij}(t)=\langle\zeta_i(t)\zeta_j(t)\rangle$:
\beq
\dot C_{jk} &=&
-2(\hat\kappa_j+\hat\kappa_k)C_{jk}
+\hat\kappa_{j-1}C_{j-1,k}
+\hat\kappa_{k-1}C_{j,k-1}
\nonumber
\\
&+&\hat\kappa_{j+1}C_{j+1,k}
+\hat\kappa_{k+1}C_{j,k+1}
+\frac{1}{2}\pi_{jk}
\eeq
where
\beq
\pi_{jk}=
\begin{cases}
\{\bar w_{j-1} +2\bar w_j +\bar w_{j+1}\}
,&k=j,
\\
-\{\bar w_{j-1}
+\bar w_j\}
,&k=j-1,
\\
0,&k<j-1,
\end{cases}
\nonumber
\eeq
and $\pi_{jk}=\pi_{kj}$. 
Taking the continuous limit we find that the source term $\pi_{ij}$ is infinitesimal:
\beq
\pi_{ij}\to 
-4\Delta x\bar Z(x)\kappa(x)\delta''(x-y)\to 0,
\nonumber
\eeq
and the remaining terms give the diffusion equation
for $C(x,y;t)\equiv
\langle\zeta(x,t)\zeta(y,t)\rangle$:
\beq
\frac{\partial C(x,y;t)}{\partial t}&=&
\frac{\partial^2(\kappa(x)C(x,y;t))}{\partial x^2}
\nonumber
\\
&+&\frac{\partial^2(\kappa(y)C(x,y;t))}{\partial y^2}.
\label{A9}
\eeq
The corresponding dynamics for the fluctuation field $\zeta$ is diffusive as well:
\beq
\frac{\partial \zeta(x,t)}{\partial t}=
\frac{\partial^2(\kappa(x)\zeta(x,t))}{\partial x^2},
\eeq
which leads to the diffusion terms to the RHS of  
the Langevin equations (\ref{Langevin}).


\begin{thebibliography}{10}

\bibitem{field98}
C.B. Field, M.J. Beherenfeld, J.T. Randerson, and P.G. Falkowski, 
Science {\bf 281}, 237 (1998)

\bibitem{franks97} 
P.J.S. Franks, 
Limnol. Oceanogr. {\bf 42}, 1297 (1997)

\bibitem{blackburn98} 
N. Blackburn, T. Fenchel, and J. Mitchell,
Science {\bf 282}, 2254 (1998)

\bibitem{martin05}
A.P. Martin,
Philos. Trans. R. Soc. London A {\bf 363}, 2663 (2005)

\bibitem{reigada03}
R. Reigada, R.M. Hillary, M.A. Bees, J.M. Sancho and F. Sagues,
Proc. R. Soc. Lond. B {\bf 270}, 875 (2003)

\bibitem{metcalfe04}
A.M. Metcalfe, T.J. Pedley and T.F. Thingstad,
J. Marine Sys. {\bf 49}, 105 (2004)

\bibitem{levy08}
M. L\`evy,
Lect. Notes Phys. {\bf 774}, 219 (2008)

\bibitem{bracco09}
A. Bracco, S. Clayton and C. Pasquero,
J. Geophys. Res. {\bf 114}, C02001 (2009)

\bibitem{mckiver11}
W.J. McKiver and Z. Neufeld,
Phys. Rev. E {\bf 83}, 016303 (2011)

\bibitem{franks97a}
P.J.S. Franks,
Limnol. Oceanogr. {\bf 42}, 1273 (1997)

\bibitem{siegenhalter93}
U. Siegenhalter and J.L. Sarmiento,
Nature {\bf 356}, 119 (1993)

\bibitem{smayda97}
T.J. Smayda,
Limnol. Oceanogr. {\bf 42}, 1137 (1997)

\bibitem{yentsch08}
C.S. Yentsch, B.E. Lapointe, N. Poulton and D.A. Phinney, 
Harmful Algae, {\bf 7}, 817 (2008)

\bibitem{assmy09}
P. Assmy and V. Smetacek,
{\it in ``Encyclopedia of Microbiology''}, edited by M. Schaechter, 
(Oxford: Elsevier, 2009), p. 27

\bibitem{scheffer97}
M. Scheffer, S. Rinaldi, Y.A. Kutznetsov and E.H. Van Nes,
Oikos {\bf 80}, 519 (1997)

\bibitem{huppert02}
A. Huppert, B. Blasius and L. Stone,
Am. Naturalist {\bf 159}, 156 (2002);
A. Huppert, B. Blasius, R. Olinky and L. Stone,
J. Theo. Biol. {\bf 236}, 276 (2005)

\bibitem{carbonel99}
C.A.A. Carbonel and J.L. Valentin,
Ecol. Model.  {\bf 116}, 135 (1999)

\bibitem{may03}
C. May, J. Koseff, L. Lucas, J. Cloern and D. Schoelhamer,
Mar. Ecol. Prog. Ser. {\bf 254}, 111 (2003)

\bibitem{smayda97a}
T.J. Smayda,
Limnol. Oceanogr. {\bf 42}, 1132 (1997)

\bibitem{huisman99}
J. Huisman, P. Van Oostveen and F.J. Weissing,
Limnol. Oceanogr. {\bf 44}, 1781 (1999)

\bibitem{cushing90}
D.H. Cushing,
Adv. Mar. Biol. {\bf 26}, 249 (1990)

\bibitem{evans85}
G.T. Evans and J.S. Parslow,
Biol. Oceanogr. {\bf 3}, 327 (1985)

\bibitem{truscott94}
J.E. Truscott and J. Brindley,
Bull. Math. Biol. {\bf 56}, 981 (1994)

\bibitem{steele81}
J.H. Steele and E.W. Henderson,
Am. Naturalist {\bf 117}, 676 (1981)

\bibitem{fasham90}
M. Fasham, H. Ducklow and S. McKelvie,
J. Marine Res. {\bf 48}, 591 (1990)

\bibitem{edwards01}
A.M. Edwards,
J. Plankton Res. {\bf 23}, 386 (2001)

\bibitem{freund06}
J.A. Freund, S. Mieruch, B. Scholze, K. Wiltshire and U. Feudel,
Ecol. Complex. {\bf 3}, 129 (2006)

\bibitem{keeling01}
M.J. Keeling, P. Rohani and B.T. Grenfell,
Physica D {\bf 148}, 317 (2001)

\bibitem{black10}
A.J. Black and A.J. McKane,
J. Theo. Biol. {\bf 267}, 85 (2010)

\bibitem{young01}
W.R. Young, A.J. Roberts and G. Stuhne,
Nature {\bf 412}, 328 (2001).

\bibitem{mckane05}
A.J. McKane and T.J. Newman,
Phys. Rev. Lett. {\bf 94}, 218102 (2005)

\bibitem{doering05}
C.R. Doering, K.V. Sargsyan and L.M. Sander,
Multiscale Modelling and Simulation {\bf 3}, 283 (2005)

\bibitem{butler09}
T. Butler and D. Reynolds,
Phys. Rev. E {\bf 79}, 032901 (2009)

\bibitem{butler11}
T. Butler and N. Goldenfeld,
Phys. Rev. E {\bf 84}, 011112 (2011)

\bibitem{siekmann08}
I. Siekmann and H. Malchow,
Math. Model. Nat. Phenom. {\bf 3}, 114 (2008)

\bibitem{holling59}
C.S. Holling,
Can. Entomol. {\bf 91}, 293 (1959);
C.S. Holling,
Can. Entomol. {\bf 91}, 385 (1959)

\bibitem{berges02}
J.A. Berges, D.E. Varela and P.J. Harrison,
Mar. Ecol. Prog. Ser. {\bf 225}, 139 (2002)

\bibitem{gonzalez08}
E.J. Gonz\'alez, T. Matsumura-Tundisi and J.G. Tundisi,
Braz. J. Biol. {\bf 68}, 69 (2008)

\bibitem{note} In order for the two species to have identical diffusion properties, 
we could assume e.g. that passive transport be dominant over swimming.

\bibitem{doi76}
M. Doi,
J. Phys. A {\bf 9}, 1465 (1976);
A.S. Mikhailov,
Phys. Lett. {\bf 85}, 214 (1981);
N. Goldenfeld, 
J. Phys. A {\bf 17}, 2807 (1985);
L. Peliti, 
J. Phys. France {\bf 46}, 1469 (1985);
H.K Janssen and U.C. Tauber,
Ann. Phys. {\bf 315}, 147 (2005)

\bibitem{vankampen}
N.G. Van Kampen,
{\it Stochastic processes in physics and chemistry}
(Elsevier, New York, 1992)

\bibitem{bonachela12}
J.A. Bonachela, M.A. Mu\~noz and S.A. Levin,
J. Stat. Phys.  {\bf 148}, 723 (2012)

\bibitem{levin76}
S.A. Levin and L.A. Segel, 
Nature {\bf 259}, 659 (1976)

\bibitem{biancalani11}
T. Biancalani, T. Galla and A.J. McKane,
Phys. Rev. E {\bf 84}, 026201 (2011)

\bibitem{montecarlo}
To this purpose, a simple algorithm evolving the fields $P$ and $Z$, at the same discrete step
as for the integration of the Langevin equation (\ref{TB2}) ($\Delta t=0.1r_0^{-1}$), was
utilized.

\bibitem{dornic05}
I. Dornic, H. Chat\'e, and M.A. Mu\~noz, 
Phys. Rev. Lett. {\bf 94}, 100601 (2005)

\bibitem{moro04}
E. Moro,
Phys. Rev. E 70, 045102(R) (2004)


\end{thebibliography}
\end{document}